# The Big Tradeoff averted: five avenues to promote efficiency and equality simultaneously

Ali Zeytoon-Nejad[1]

## Abstract
Society as a whole faces a host of economic tradeoffs, many of which emerge around economic policies. An example of tradeoffs that any society faces in many economic realms is the tradeoff between economic efficiency and income equality (aka the efficiency-equality tradeoff). This tradeoff has been called "the Big Tradeoff" by the esteemed economist Arthur Okun, who also termed it "the Double Standard of a Capitalist Democracy." Although the efficiency-equality tradeoff is more or less an inevitable tradeoff in most societal settings and economic contexts, there are still some special circumstances in which this tradeoff can be avoided. This paper identifies five such avenues and elaborates on why and how the tradeoff between these two somewhat contradictory societal goals—efficiency and equality—can be deftly averted under the mentioned circumstances. These avenues with their transformative potential can and should be used so that a capitalist society as an integrated whole can promote both efficiency and equality at the same time under these scenarios and avoid facing the Big Tradeoff in cases where it is evitable. Static and dynamic economic models are developed, solved, and applied to facilitate the articulation and exposition of the main points of each solution with formal rigor and logical coherence. Finally, policy implications are discussed.

**JEL Classification** A1 · A13 · D6 · D61 · D63

---

✉ Ali Zeytoon-Nejad
zeytoosa@wfu.edu

[1] School of Business, Wake Forest University, 1834 Wake Forest Rd, Farrell Hall, Building 60, Winston-Salem, NC 27109, USA

## 1 Introduction

All decisions involve tradeoffs. Not only humans at an individual level but also society as a whole faces tradeoffs. Such tradeoffs arise very often when making decisions about economic policies. A famous example of economic tradeoffs that any society faces is the tradeoff between "economic efficiency" and "income equality," which is also known as the efficiency-equality tradeoff. Efficiency refers to the state of the world in which an economy's resources are allocated in such an optimal way that maximizes prosperity and benefits, and minimizes costs and waste, resulting in the highest amount of income and living standards for the average member of the society, given the available amount of resources and the available level of technology in that economy. In contrast, equality refers to how equally that prosperity (i.e., efficiency) is distributed among the society's members or different social groups in that society. The prominent American economist Arthur Okun (1975)[1] aptly labeled this delicate balance between economic efficiency and income equality as "the Big Tradeoff." In his astute observations as the Chairman of the Council of Economic Advisers (CEA) to the US President, he further elucidated this tradeoff as "the Double Standard of a Capitalist Democracy" that it professes and pursues an egalitarian political and social system while simultaneously gapes disparities in the economic well-being of its citizens. He goes on to say that "to the extent that the [economic] system succeeds, it generates an efficient economy. But that pursuit of efficiency necessarily creates inequalities. And hence society faces a tradeoff between Equality and Efficiency." Okun cogently posits that while the pursuit of an efficient economy often yields favorable outcomes, it inevitably gives rise to inequalities.[2] Consequently, society as a whole finds itself at a crossroad, confronting the fundamental tradeoff between equality and efficiency.

Efficiency is obviously a desired outcome, as getting more from a scarce resource is certainly better than getting less *ceteris paribus*. That is, a higher standard of living is undoubtedly better than a lower standard of living. Efficiency takes place when firms reduce their costs and increase their output and trade in the market, and consumers enjoy lower prices for goods and services they buy and also get to engage in more mutually beneficial trades with firms. Equality can also be a desired outcome, as it can promote social cohesion, improve the sense of happiness and satisfaction within society, and at the same time, lower crime rates, and reduce political instability, as shown and discussed by a multitude of studies.[3] Thus, both of these economic outcomes are desired outcomes, and an ideal society would aim for the levels of both of these two outcomes to be as high as possible in the society; however, it is an unfortunate fact that there usually exists a tradeoff between these two appealing ends, meaning that when the government aims to increase one (e.g., when the government aims to increase equality through taxing the rich and redistributing it to the poor), the other one tends to decrease (e.g., the rich will lose their incentives to invest and produce in the regions where they are taxed heavily, so they may choose to engage less in economic activity or leave that region, resulting in decreased economic activity and lowered economic efficiency in that region).

---

[1] This book was initially written by Okun (1975) and was re-published in recent years (with some additions from Okun's other works) under Okun and Summers (2015) with a foreword written by Lawrence Summers, another renowned American economist.

[2] Zeytoon-Nejad (2024) has conducted a comparative study of 135 countries using a panel-data approach and has shown this empirical fact. According to his empirical findings, as countries' forms of economic systems become closer to that of the free-market capitalism, their levels of income per capita (as a measure of efficiency) increases while their levels of income equality tend to decrease. As shown in Zeytoon-Nejad (2024), this tradeoff relationship is robust and holds true regardless of whether the Gini measure of market-income inequality or the Gini measure of disposable-income inequality is considered.

Although the efficiency-equality tradeoff is more or less an inevitable tradeoff in most socio-economic contexts, there are still certain circumstances in which this tradeoff can be averted. This paper identifies and examines five such avenues and elaborates on the reasons and methods through which the tradeoff between the two societal goals of efficiency and equality can be adroitly averted under the said circumstances. These avenues can and should be used so that a capitalist society as a whole can promote both efficiency and equality at the same time under these scenarios and evade the significant challenges posed by the so-called "Big Tradeoff" at least in cases where it is possible to do so. These avenues are indeed five solutions that offer potential means to somewhat address the prevalent efficiency-equality tradeoff that tends to arise under capitalist systems. The five avenues to be introduced and discussed in this paper are as follows.

- Utilizing expansionary fiscal policy when aggregate demand (AD) falls short of aggregate supply (AS) (i.e., AD < AS): Stimulating AD by targeting those with the highest Marginal Propensity to Consume (MPC)
- Providing equal opportunity for empowerment for the needy: Short-run public pain, long-run public gain
- Optimizing the timing of raising the minimum wage in the economy to promote both efficiency and equality simultaneously: Raising the minimum wage when the actual rate of unemployment is below the natural rate of unemployment
- Motivating social cooperation in the case of externalities: Equality and efficiency in the usage of natural resources
- Encouraging charitable engagement for social impact in the economy: Maximizing the economic pie for the poor and maximizing the utility pie for the rich

---

[3] Examples of such studies include Alesina & Perotti (1996), Kawachi et al. (1999), Wilkinson & Pickett (2011), Helliwell & Huang (2014), McCarty et al. (2003), Duca and Saving (2016), and Gu and Wang (2022). Wilkinson and Pickett (2011) argue that greater income equality can contribute to stronger social cohesion, increased happiness, lower crime rates, and enhanced political stability. According to Kawachi, et al. (1999), income inequality is associated with higher rates of crime. Alesina and Perotti (1996) report that lower income inequality is associated with reduced political instability. Helliwell and Huang (2014) find that more equal societies tend to have greater levels of subjective happiness and life satisfaction. McCarty et al. (2003), Duca and Saving (2016), and Gu and Wang (2022) discovered a positive association between income inequality and political polarization, implying that higher levels of income inequality are linked to increased political divisions within societies.

In the remainder of this paper, for the aforementioned items, static and dynamic economic models will be developed, solved, and employed in order to ensure a formal and logical presentation of each solution. Afterwards, a comprehensive conclusion will be derived from the overall discussion, and the key findings are summarized.

## 2 Main discussion

The literature on the tradeoff between income equality and economic efficiency offers a nuanced understanding of the interplay between these societal goals, which highlights the necessity of considering these dynamics in the formulation of socio-economic policies. Numerous recent studies have contributed to this discourse by emphasizing the intricate relationship between redistribution efforts and income equality on the one hand and economic efficiency and growth on the other hand. Woo's (2020) empirical investigation provides a comprehensive analysis spanning several decades and a multitude of countries, revealing that there is a non-trivial tradeoff between income equality and economic efficiency. By systematically examining the impact of redistribution on economic performance, Woo (2020) underscores the complex nature of policy interventions aimed at addressing income inequality. Simões et al. (2013) offers a regional perspective on inequality and economic growth in Portugal. Their findings reveal a negative impact of inequality on per-capita output at the lower end of the earnings distribution (likely due to its dampening effect on investments in human capital) and a positive impact of inequality on per-capita output at the top end of the distribution (supporting the incentives argument in the inequality-growth nexus).

Andersen and Maibom (2020) shed light on this tradeoff relationship particularly within OECD countries. Their findings suggest that, while increased redistribution can promote equality, it often comes at the cost of reduced economic efficiency, as reflected in a larger tax burden associated with lower efficiency and heightened equality. In fact, consistent with standard economic theory, they conclude that a larger tax burden is associated with lower economic efficiency and more income equality. Andersen et al. (2021) scrutinize pension reform impacts on the efficiency-equality tradeoff through an investigation of a Norwegian early retirement program reform and unveils heightened old-age labor supply alongside increased income inequality. They conclude that, despite elevated hours worked, the minimal impact on inequality across income levels in their study underscores the need for an intricate balance between income equality and economic efficiency in pension system reforms.

In addition, there have been studies in the literature that offer insights into potential mitigating factors that could attenuate the tradeoff between income equality and economic efficiency. Røed and Strøm (2002) argue that in certain contexts, progressive tax policies may not necessarily hinder efficiency. Instead, in non-competitive labor markets, progressive taxes might incentivize wage moderation, leading to a more efficient allocation of resources. This perspective somehow challenges the notion of an inherent tradeoff between equity and efficiency, suggesting that under specific conditions, redistributive measures can be designed to enhance both objectives simultaneously. Moreover, Atkinson (2015) provides an examination of the extent, origins, and changes in income inequality on a global scale. Thorbecke (2016) critically reviews Atkinson's book and addresses the issue of the tradeoff between efficiency and equality, stating that it is a crucial aspect of any growth or development strategy. Indeed, if there is a tradeoff between these two goals, any effort to reduce inequality would entail a potential decrease in efficiency. However, as Atkinson concludes, under specific circumstances, there may be no such tradeoff, and within certain limits, reducing inequality could actually enhance efficiency. This suggests the possibility of achieving both more equality and more efficiency simultaneously. As Atkinson articulates, "Equity and efficiency may align in the same direction," implying that there could be no tradeoff between the two under certain circumstances. This underscores the importance of adopting a nuanced approach to policy-making, one that recognizes the potential for synergy between equality and efficiency objectives.

Motivated by these diverse perspectives within the respective literature, the present paper aims to contribute to this discourse by introducing five circumstances under which income equality and economic efficiency can be promoted simultaneously. Building upon the insights gained from prior research, the present paper seeks to identify potential avenues for policymakers to navigate the complex landscape of socio-economic policy-making. In fact, this section of the present paper introduces, explains, and exemplifies the five aforementioned avenues through which the inconvenient tradeoff relationship between efficiency and equality can be avoided in the realm of economic policy. It provides a detailed exploration of how policymakers can navigate this intricate relationship by strategically designing, formulating, and implementing economic policies. By delving into these avenues, policymakers can use innovative approaches to align and harmonize efficiency and equality objectives while ensuring effective policy outcomes as well.

### 2.1 Utilizing expansionary fiscal policy when aggregate demand (AD) falls short of aggregate supply (AS) (i.e., AD < AS): stimulating aggregate demand by targeting social groups with the highest Marginal Propensity to Consume (MPC)

Fiscal policy is defined as the use of government spending and tax policies to influence a country's economy mainly through adjusting AD in order to promote economic growth and stabilize the macroeconomic environment over the course of business cycles. As such, the major tools of fiscal policy include government spending (increases and cuts) and taxes (increases and cuts). In a sense, the primary mission of fiscal policy, as proposed by John Maynard Keynes, is to prevent recessions and inflation from occurring in the economy. Figure 1 exhibits two types of gaps between AD and AS called "recessionary gaps" and "inflationary gaps," which arise from discrepancies between AD and AS in the economy throughout business cycles. From an efficiency point of view, these gaps reflect the underutilization or overutilization of resources in an economy.

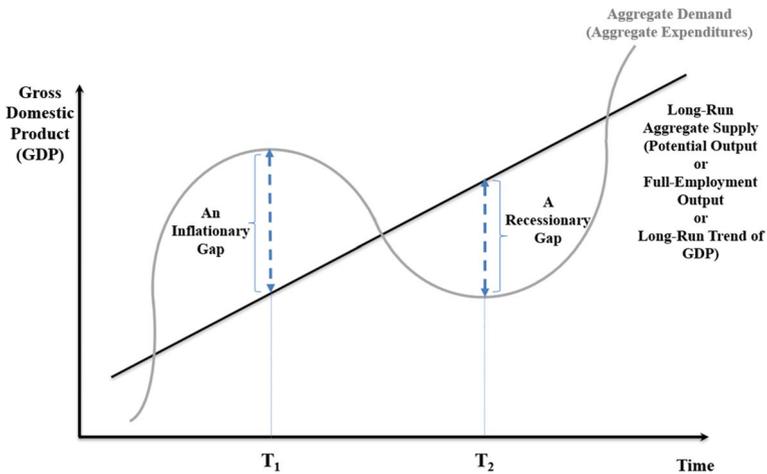

**Fig. 1** Aggregate demand, aggregate full-employment supply, recessionary gaps, and inflationary gaps in a dynamic model. Note: Aggregate expenditure (which forms AD) is the total amount spent for the economy's output. When aggregate expenditure (AD) falls below aggregate output (AS) (like what has occurred at time $T_2$ in Fig. 1), inventories will increase, and the economy will experience a recessionary gap, which leads to a rise in unemployment. In contrast, when aggregate expenditure (AD) exceeds aggregate output (AS) (like what has happened at time $T_1$ in Fig. 1), inventories will fall below the equilibrium amounts of inventories, and the increased demand leads to an inflationary gap, during which prices increase due to demand-pull inflation

Figure 2 shows the historical trend of aggregate demand and output (i.e., Gross Domestic Product, or for short GDP) in the US for the time period of 1947 until 2022. According to the schematic, dynamic AS-AD model provided above in Fig. 1, during the periods where AD in the US (i.e., the blue line in Fig. 2) falls above AS in the US (i.e., the red line in Fig. 2), the economy tends to experience periods of inflation (like many years during the 1960s and 1970s in the US, as shown below in Fig. 2). According to the schematic, dynamic AS-AD model provided above in Fig. 1, during the periods where AD in the US (i.e., the blue line in Fig. 2) falls below AS in the US (i.e., the red line in Fig. 2), the economy tends to experience periods of recessions (like during 2007–2008 in the US, as shown below in Fig. 2).

Figure 3 shows the correspondence between the dynamic model of output gaps (as demonstrated in Fig. 1) with the conventional, static AS-AD model of output gaps in economics.

Put simply, recessions are defined as declines in the size of economic activity. As such, recessions have immediate negative effects on economic efficiency. This is because AD falls short of AS during recessions, meaning that the economy's demand for goods and services falls below the economy's potential (full-employment) supply capacity. This is clearly a case of inefficiency, where there is a waste of resources in the economy (e.g., idle factories, unutilized or underutilized productive capacity, and unemployed labor). Similarly, inflation can have negative effects on the economy's prosperity, especially when the rate of inflation exceeds the commonly

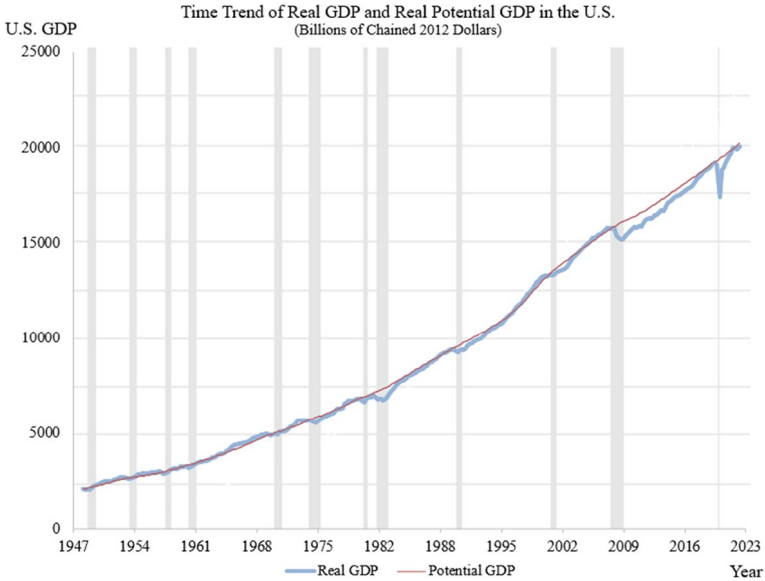

**Fig. 2** The actual time trend of real GDP and real potential GDP in the US. Source of data: Bureau of Economic Analysis (BEA); Congressional Budget Office (CBO)–FRED Dataset. Note: Shaded areas indicate U.S. recessions

targeted rates of inflation set by central banks. This may result in negative effects on economic efficiency (e.g., through the destabilization of the macroeconomy, elevated levels of uncertainty in the economy, risk-averse agents operating in a volatile

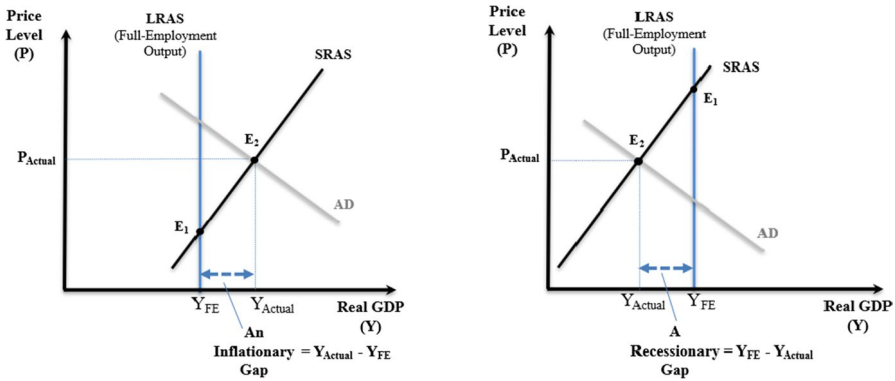

**Fig. 3** Aggregate demand, aggregate supply (both short-run and long-run), recessionary gaps, and inflationary gaps in a static AS-AD model. Note: The left-hand-side diagram depicts the corresponding AS-AD situation to Fig. 1 at time $T_1$ from a static perspective. The gap between the $Y_{Actual}$ and $Y_{FE}$ is an inflationary gap of size ($Y_{Actual} - Y_{FE}$). On the contrary, the right-hand-side diagram depicts the corresponding situation to Fig. 1 at time $T_2$ from a static perspective. The gap between the $Y_{FE}$ and $Y_{Actual}$ is a recessionary gap of size ($Y_{FE} - Y_{Actual}$). The SRAS curves represent short-run AS curves, which explain short-run fluctuations in the economy. Additionally, note that, for the sake of clarity, in both diagrams, the initial AD curves are assumed to be located on the two $E_1$ points, but they are not depicted in these diagrams. This omission aims to avoid overcrowding the diagrams and maintain readability

economy experiencing a utility loss, formation of inflationary expectations and the threat of stagflation, and temporal business decisions ending up being suboptimal *ex-post* due to the high degrees of fluctuations in markets, all of which can cause inefficiencies). As such, it can be said that the broad mission of fiscal policy (and monetary policy) is to preserve economic efficiency primarily through economic-stabilization policies.

Although there usually exists a tradeoff relationship between efficiency and equality in many economic contexts, there are circumstances and ways in which fiscal policy can be used not only to promote efficiency but also to promote equality at the same time. One such instance is when AD falls short of AS in an economy, and the government intends to stimulate the economy through an expansionary fiscal policy and, in particular, through a tax cut. Under such circumstances, if the government designs the tax cuts to benefit only the poor (or at least benefit the poor relatively more than it benefits other social groups), then not only the efficiency purpose of the policy (which is to increase AD to meet AS soon) is achieved more effectively and more quickly, but it also promotes equality within the economy concurrently. For example, two approaches that can be taken to implementing such a tax-cut policy design are as follows: Expanding the tax bracket of the zero-tax rate (also known as the "tax-exempt bracket" or the "non-taxable income bracket") in order to encompass a broader segment of the impoverished population, and/or reducing the marginal tax rates of the subsequent tax brackets following the tax-exempt bracket (which apply to segments of the population with low levels of income), while keeping those of the other social groups unchanged.

The reason why the efficiency objective of the policy is achieved more effectively and more quickly is because of the fact that the poor tend to have a higher Marginal Propensity to Consume (MPC)[4] than the rich, so they are very likely to spend all or a very large portion of the proceeds of the tax cuts on consumption, which leads to a substantial increase in AD, which helps in accomplishing the mission of the fiscal policy. Thereby, the poor's increased consumption will effectually increase AD, which in turn helps more effectively in stimulating the economy, so that the economy recovers faster and returns to its full-employment level more quickly and more effectively for a given amount of fiscal stimulus provided by the government. This form of conducting fiscal policy would help the economy to increase efficiency, but at the same time, help the economy promote equality, too, because the poor are eventually receiving a higher amount of disposable income, and thereby the income equality in that economy will improve *ceteris paribus*.

---

[4] The Marginal Propensity to Consume (MPC) is defined as the increase in a consumer's spending after an increase occurs in their disposable income (i.e., income after paying taxes and receiving transfers). As such, it is mathematically defined as MPC = $\frac{\Delta C}{\Delta Y^d}$, where $\Delta C$ denotes the change in consumption spending and $\Delta Y^d$ denotes the change in disposable income. Put differently, the proportion of disposable income that an individual spends on consumption is called propensity to consume (for short, $PC = \frac{C}{Y^d}$), and the proportion of one additional unit of disposable income that an individual decides to consume is called "marginal" propensity to consume (for short, MPC = $\frac{\Delta C}{\Delta Y^d}$). For instance, when an individual's disposable income goes up by $1, the individual consumes 80 cents of the additional income, then that individual's MPC will be equal to 0.80. By definition, MPC always falls between the lower bound of 0 and the upper bound of 1, because when we rule out the possibility of borrowing, the individual cannot increase their consumption spending more than the additional disposable income that they have received. The MPC tends to be higher and **closer to 1 for the poor** (because the poor have many basic needs that are not fulfilled yet). The MPC tends to be relatively lower and **closer to 0 for the rich** (because their basic needs are more likely to have been met already, so, when there is a tax cut, for example, they tend to save the whole or at least a sizable portion of the proceeds of the tax cut, which is their marginal disposable income in this case). To see some empirical estimates of the MPC of different income groups, you can see Carroll et al. (2017).

One caveat that is noteworthy with respect to the conduct of this type of expansionary fiscal policy is that policy makers must consider a priori the effect of differential MPCs of different income groups on the scope and the effectiveness of the policy. For a given recessionary gap, the total amount of the fiscal stimulus package needed to be provided to the poor is always smaller than the total amount of the fiscal stimulus package to be given to the rich (of the same population size) to achieve the same stabilizing goal. This is because the poor would consume a relatively larger portion of the whole fiscal stimulus package given to them compared to that same amount given to the rich (who would get to save part of the proceeds), so a smaller size of fiscal stimulus package to the poor would have the same effect as a larger fiscal stimulus package given to the rich. As an empirical example, Carroll et al. (2017) have estimated that a fiscal stimulus that is targeted toward income groups in the bottom half of the wealth distribution in the US would result in 2 to 3 times more effectiveness in increasing AD than just a blanket stimulus that targets all income groups similarly.

Therefore, in order for the government to ensure that they do not overdo this fiscal policy, they need to carefully consider the size of the stimulus package that is provided to each particular income group proportionately to the respective income group's aggregate MPC values. If policy makers implement this type of expansionary fiscal policy under the false assumption that all income groups have the same MPC and disregard the fact that the poor and the middle-class have higher MPCs than the rich, then they are subject to over-stimulating the economy (by injecting an overly large amount of stimulus into the economy than what is needed to just return the economy to its full-employment output). This incident seems to have occurred in the US economy after distributing Covid-19-related fiscal stimulus packages which were provided mostly to the poor and the middle-class, probably disregarding the effectiveness of such fiscal stimuli in increasing AD when they are offered to the lower-income and middle-income groups, which in turn overly heated the economy, resulting in an inflationary output gap and high inflation rates that took place in 2021 and 2022 possibly partly due to the overdoing of this policy.[5] To conclude, when in a recessionary gap, targeting the poor in the design of an expansionary fiscal policy would help more effectively the economy in returning it back to the potential output

---

[5] This is only one potential reason for the high rates of inflation that occurred in the US economy during 2021 and 2022. It can be argued that a multitude of reasons contributed to the formation of the mentioned inflationary period in the US. These include, but certainly are not limited to, (1) COVID-related issues that had negative effects on real economic activity (e.g., an inevitable shortage of labor during COVID, labor's delayed return to work during the aftermath of COVID, and domestic supply-chain disruptions and issues), (2) the elevated prices of oil even before the Russian invasion of Ukraine, (3) Russian-invasion-related issues (e.g., a supply shock due to quickly increased oil prices, supply-chain disruptions and issues related to the Russian invasion such as grains' supply-chain issues and elevated

trend.[6] The implementation of such an expansionary fiscal policy would improve both equality and efficiency at the same time.

## 2.2 Providing equal opportunity for empowerment for the needy: short-run public pain, long-run public gain

Providing equal opportunities and implementing empowerment initiatives for the needy in order to enable them to become economically independent members of society may seem to be a pressure on the public sector's budget in the short run, but it definitely comes with long-run gains for society as a whole and generates considerable cost-savings for the public sector's budget in the long run, when compared to the counterfactual. Such empowerment initiatives can serve as a strategy that aims to promote both equality and efficiency in society.

It is easy to see how such empowerment programs can promote equality in society. Take education as an example. This approach helps in reducing societal disparities by offering resources to those in need for their education (which promotes the equality of income in society) and also by providing equal empowerment opportunities to those in need, such as access to quality education (which promotes the equality of opportunity). Such empowerment programs can also result in the promotion of efficiency in society. By empowering marginalized individuals and communities, society can tap into a wider pool of productive members with wider range of talents, skills, and perspectives, which can result in fostering human capital, innovation, productivity, and economic growth in the society, provided that such empowerment efforts are made in the direction of the society's actual needs.

Student financial assistance programs (also known as student loan programs) are indeed a compelling example of empowerment initiatives aimed at fostering equal opportunity, promoting efficiency, and generating positive externalities. This integrated approach serves as a comprehensive solution to achieve all these three objectives effectively and cohesively. By prioritizing accessible and high-quality higher education for all talented individuals, regardless of their financial means,

---

Footnote 5 (continued)

food prices), (4) the delayed conduct of contractionary monetary policy by the monetary authority (i.e., the Fed) in the US and not precisely accounting for the time lags in the conduct of monetary policy and its effects on the economy and starting to raise interest rates somewhat late and at a low pace, (5) the unfortunate overlapping of COVID fiscal stimuli and presidential-election campaigns, which somehow resulted in an unfortunate blending of economic policy and political motives involved around the election to increase the chance of winning the election by one party and to keep the promises made during the campaign by another party, and (6) the fiscal authority possibly not accounting beforehand accurately for the different effects of differential MPCs of different income groups on the scope and the effectiveness of the policy, as elaborated in the main body of the paper.

[6] On the contrary, when in an inflationary gap, targeting a reduction in the spending of the people in lower-income groups in the design of contractionary fiscal policy helps more effectively the economy in returning it back to the potential output trend. Although such a policy move increases efficiency more effectively and more quickly, it may result in lowering equality, too. In that situation, there will arise an inevitable tradeoff between efficiency and equality.

governments can level the playing field and ensure that every talented student has a fair chance to succeed.

When individuals lack access to quality education and equal opportunities, they may face greater challenges in their personal and professional lives. This can result in increased reliance on social welfare programs, higher unemployment rates, and reduced productivity. With the passage of time, the cumulative effects of these events can overload the economic system and impose a significant financial burden on the public sector's budget and society as a whole, which necessitates substantial resources to address the resulting social and economic issues, while they could have been avoided if those individuals needing financial means to become financially independent had been provided with the tools and resources at the right time when they needed those resources. Societies can effectively mitigate such potential welfare-based future costs by providing talented individuals who are likely to become self-sufficient, contribute to the economy, and reduce their reliance on public assistance in the long run.

In addition to cost-savings in welfare programs, programs supporting higher education carry multiple additional benefits that extend beyond individual outcomes. When talented individuals from financially disadvantaged families are given equal access to higher education, it not only improves their personal prospects but also creates positive externalities that contribute to the betterment of society as a whole in several ways, including, but certainly not limited to, more value-added generated in the economy, reduced rates of crime, and more informed voters to make well-informed civic decisions.

A rational way of understanding whether it is reasonable (from an efficiency point of view) for the public sector to engage in the provision of empowerment programs such as financial assistance programs for students in higher education is to conduct a rational, marginal cost–benefit analysis for such a program from the viewpoint of the government's public budget. Assume that the government is currently considering whether it should provide subsidized loans to talented students in need of loans (as a nudge) for them to attend universities. For the sake of simplification, assume that a typical student enters higher education at the beginning of their 20th year of life ($t=0$); graduate after 4 years ($t=4$), start working, earning, and repaying their student loan a year after ($t=5$); continue to repay their student loan for 25 years until the age of 49 ($t=29$); work until the age of 65 ($t=45$) and then retire; and live until the age of 77 ($t=57$), which is the current life expectancy in the U.S. Note that this rational dynamic, marginal cost–benefit analysis is conducted from the perspective of the government which, as a benevolent social planner, is considering whether it is publicly beneficial to provide students with subsidized loans that work as a nudge to attend university for those qualified students who marginally need the loans to be able to attend school. From an efficiency point of view, such a program is economically justified for the public sector to provide if the following condition holds:

$$\text{NPV(Marginal Social Benefits)} > \text{NPV(Marginal Social Costs)} \qquad (1)$$

As noted earlier, such an empowerment program can have at least three benefits for society, including more value-added generated in the economy due to increased human capital, cost savings in welfare programs, and the gains of positive externalities. Accordingly, the net present value (NPV) of the marginal social benefit of such an empowerment program will be equal to:

$$\text{NPV(marginal benefits)} = \sum_{t=5}^{45} \frac{HCVA_t}{(1+i_{social})^t} + \sum_{t=5}^{57} \frac{WPCS_t}{(1+i_{social})^t} + \sum_{t=5}^{57} \frac{PEG_t}{(1+i_{social})^t} \tag{2}$$

where $HCVA_t$ denotes the value added to the economic pie in period $t$ due to the acquired human capital through education, $WPCS_t$ denotes the cost savings in welfare programs in the future, $PEG_t$ denotes social gains derived from positive externalities, $i_{social}$ denotes the social discount rate (a measure used to evaluate the present value of future benefits for public-sector projects), and $t$ denotes the time period.

To accurately assess the marginal cost to the government of student loans, careful consideration must be given to two key cost components, including the principal amount disbursed to students as loans and the time value of that money, represented by the interest accrued on the principal. That time value of money for the government aligns with its cost of borrowing, which can be considered to be the free-market equilibrium interest rate (that is, the rate at which a government with a balanced budget gets to borrow funds in the market for loanable funds). However, the government makes student loans available to students at a subsidized interest rate, which is typically lower than the free-market equilibrium interest rate.

In a rational marginal cost–benefit analysis, the principal amount and the portion of the free-market interest rate that is repaid by the student in the future become irrelevant, since the government provides the principals and subsequently recovers them along with the subsidized interests on the loans. Therefore, the only marginal cost to the government arising from the student loan program is the difference between the free-market equilibrium interest rate (denoted by $r_{FM}$) and the interest rate charged to the students for the subsidized loans (denoted by $r_{SL}$).[7] Then, the net interest rate that the government has to incur due to offering this student loan program is equal to $\Delta r = r_{FM} - r_{SL}$,[8] which is the effective basis for computing the marginal cost of the student loan program. When applied to the principal of the loan (denoted by $L$) which is spread over the loan's duration as annual repayment cash flows and when multiplied by the respective remaining balance of the loan ($LB_t$) in each period, the sum of the present values of all the loan repayments adjusted by the difference interest rate ($\Delta r$) will represent the marginal cost of this student loan program. In fact, each (annual) Loan Repayment Cash Flow based on $r_{SL}$ (denoted by $LRCF_t^{SL}$) needs to be adjusted to account for the difference interest rate ($\Delta r$), and the resulting adjusted Loan Repayment Cash Flow based on $r_{FM}$ (denoted by $LRCF_t^{FM}$) will be equal to $LRCF_t^{FM} = LRCF_t^{SL} + (\Delta r \cdot LB_t)$. Thus, the relevant cost to consider as the marginal cost emerges from the differences between $LRCF_t^{FM}$ and $LRCF_t^{SL}$, which can be denoted by $\Delta LRCF_t = LRCF_t^{FM} - LRCF_t^{SL}$.[9] Then, the following expression will yield the overall marginal cost of offering this student loan program for the government.[10]

---

[7] It is important to note that another cost component of providing such a public program for the government is the administrative cost of running this public program. However, in the context of the calculations conducted in this section, this administrative cost has been omitted due to its relatively smaller magnitude compared to the computed marginal cost mentioned above.

[8] This rate represents the marginal (additional) interest that the government could have earned in this loan program if it had charged the free-market rate to the student borrowers. In other words, it is the potential earnings that the government has foregone due to the lower (subsidized) interest rate that it charges to the students.

$$\text{NPV(marginal costs)} = \sum_{t=0}^{29} \frac{\Delta LRCF_t}{(1+i_{social})^t} \quad (3)$$

Accordingly, the evaluation of the marginal costs and benefits of the student loan program from the government's perspective can be summarized as follows:

$$\sum_{t=5}^{45} \frac{HCVA_t}{(1+i_{social})^t} + \sum_{t=5}^{57} \frac{WPCS_t}{(1+i_{social})^t} + \sum_{t=5}^{57} \frac{PEG_t}{(1+i_{social})^t} > \sum_{t=0}^{29} \frac{\Delta LRCF_t}{(1+i_{social})^t} \quad (4)$$

Based on three reasons discussed below, it is highly plausible that, for a reasonably-sized subsidized interest rate, the net present value (NPV) of the marginal benefits of this program for the public sector considerably outweighs the NPV of its marginal costs. That is,

$$\text{NPV(marginal benefits)} > \text{NPV(marginal costs)} \quad (5)$$

There are three reasons supporting the assertion that the benefits of such a student loan program are likely to outweigh the costs associated with it for the public sector. Firstly, the duration of costs is typically much shorter compared to the longer duration of future benefits, resulting in a higher value on the left-hand side of the above inequality. Secondly, there are multiple channels through which society can benefit from such programs, while there is only one major cost component for the program from the viewpoint of the public sector, which is limited to a small fraction of the potential interest rate that the government has to forgo, as previously explained. Thirdly, the typically low social discount rate that is applicable to public programs of this nature assigns greater weight to future benefits compared to when the discount rate is larger. This is because a typically low (social) discount rate discounts future benefits less heavily than a higher discount rate would. Therefore, economically, there is a clear justification in favor of this type of empowerment program, as their marginal benefits are very likely to exceed their marginal cost for the key reasons mentioned above.

---

[9] The total marginal cost of the program can be expressed in several equivalent ways. For example, note that $LRCF_t^{FM} = LRCF_t^{SL} + \Delta r \cdot LB_t$. Then, $LRCF_t^{FM} - LRCF_t^{SL} = \Delta r \cdot LB_t$. Accordingly, the overall marginal cost of this program for the government can be written also as $\sum_{t=0}^{29} \frac{\Delta r \cdot LB_t}{(1+i_{social})^t}$. These two representations are inherently equivalent, and the only apparent difference lies in their formulations.

[10] It is important to note that this calculation considers the difference in interest rates as the primary factor impacting the overall marginal cost. However, other factors such as default rates, administrative costs, and other loan fees may also affect the overall marginal cost to the government. Additionally, one may argue that empowering the needy or student loans can have associated negative impacts and externalities on an economy from the perspective of public economic theory, meaning that these policies may lead to unintended negative consequences for society as a whole and have adverse effects on the broader economy. For example, one may argue that student loans can lead to high levels of debt among graduates, potentially reducing their ability to participate in the economy by delaying major purchases such as homes or cars. Additionally, if a significant portion of borrowers default on their loans, it could strain government budgets or financial institutions, leading to broader economic repercussions. Similarly, needy-empowerment programs, if not carefully designed, may create disincentives for individuals to seek employment or engage in productive activities, which ultimately may reduce overall economic output. Moreover, these programs could contribute to dependency and perpetuate cycles of poverty rather than promoting self-dependence and economic growth. Thus, these are additional aspects and potential consequences of such policies which should be considered in their designs to ensure that such possibilities are either ruled out or the likelihood of these possibilities is minimized.

All in all, the analysis reveals that the NPV of the future benefits of this type of empowerment program outweighs the NPV of the public resources allocated towards such an empowerment program. This indicates the importance of implementing timely initiatives today in society in order to mitigate the need for larger expenditures in the future to provide long-term, permanent assistance to these groups. Empowerment initiatives of this nature can serve indeed as a catalyst to transform talented individuals, who may otherwise cause financial burdens on the public budget, into productive, independent members of society. By providing them with the necessary support and opportunities at the right time, these initiatives enable individuals to contribute to the economy and achieve financial independence. While there may be short-term costs involved for the public sector, the longer-term values to be generated and cost-savings to be realized justify the short-term, smaller costs. This approach helps society evade potential long-term challenges that would arise in the counterfactual, alternative state of the world in the absence of such initiatives. This type of empowerment program serves as an exemplary effort that embodies not only the dual objectives of promoting income equality and promoting economic efficiency, but also providing equal opportunities and generating positive externalities.

After all, it is important to recognize that the examination conducted above primarily focuses on the instrumental value of such initiative. However, it is also important to take into account the intrinsic value of offering such help to individuals in need, implementing empowerment initiatives for them, and providing equal opportunity for all, as every individual deserves the chance to fulfill their potential and lead a dignified life. The intrinsic value of such programs upholds the core principles of fairness in equal opportunity and human dignity. Implementing empowerment programs of this nature align not only with the ideals of both efficiency and equality, but they also create a more inclusive, fair, meritocratic, and prosperous society for all.

### 2.3 Optimizing the timing of raising the minimum wage in the economy to promote both efficiency and equality simultaneously: raising the minimum wage when the actual rate of unemployment is below the natural rate of unemployment

The minimum wage is a type of price control policy referred to as a price floor. A minimum-wage law is binding and effective if the minimum wage is set above the free-market equilibrium wage for the respective group of labor. Accordingly, employed individuals who are paid wages below the officially mandated minimum wage have the right to legally challenge their employers. In general, this type of policy is formulated with the ultimate intention of promoting equality in favor of those in need. Inherent in this type of policy is the implicit assumption that the people working for the minimum wage are the needy, and that a sort of transfer of income from employers to employees will take place as a result of this law from those who are relatively wealthier to those who are relatively poorer. One major drawback of minimum wage laws is the structural unemployment that they generate in the economy, since some employers may be unable or unwilling to hire workers at a higher wage rate than the free-market equilibrium wage. As a consequence, it causes loss of welfare for the market participants in the labor market, generating a deadweight loss in the labor market, and thereby reducing efficiency in the economy. In fact, although the policy maker initially aims for higher equality in society, society ends up with less efficiency ultimately.

While the initial intention of promoting equality in this case is commendable, the unintended consequence is a tradeoff between equality and efficiency. However, this paper argues that this tradeoff can be avoided at least in circumstances where the actual rate of unemployment falls below the natural rate of unemployment.[11] This proposal will be elaborated on in what follows. Before that, it is helpful to analyze the historical time trend of the minimum wage in the U.S. Figure 4 displays the temporal evolution of both the nominal and real federal minimum wage in the US over the past few decades.[12]

The stepwise blue line depicted in this diagram represents the nominal minimum wage enacted by Congress and signed into law by presidents from 1949 until 2022. This line shows that the minimum wage in the US has gradually increased from $0.40 per hour in 1949 to $7.25 per hour in 2022. The red line in this diagram represents the real minimum wage during the same period of time. This real measure of the minimum wage indicates the real purchasing power of the minimum wage by excluding the effect of inflation occurring in the economy. Specifically, the red line represents the purchasing power of the (nominal) minimum wage in terms of the average purchasing power of dollars during the period of 1982–1984.

Fluctuations in the red line indicate two events. Sharp increases in the line represent increases in the (nominal) minimum wage done by Congress (each of which correspond with a step in the staircase blue line, but the red line is adjusted for being in terms of real purchasing power). Gradual decreases in the red line represent the reductions in the purchasing power of the newly set minimum wages over time due to inflation. As evident in this diagram, although the nominal minimum wage has constantly been increasing, the real minimum wage has almost always remained in the corridor of $6 to $10 with an average value close to $8.20 in terms of the average purchasing power of dollars during the period of 1982–1984. Table 1 reports several summary statistics of these two variables in order to provide more detailed information about the dynamics of the changes in the minimum wage.

---

[11] Another term that is commonly used to refer to the natural rate of unemployment is NIRU, which stands for Non-Accelerating Inflation Rate of Unemployment. The natural rate of unemployment refers to the lowest level of unemployment that an economy can sustain without triggering inflationary pressures. In fact, this rate represents the equilibrium unemployment rate in which the labor market is in balance, and there is no upward or downward pressure on wages that could lead to inflation or deflation in the economy.

[12] It is important to note that this graph shows the federal minimum wage. However, individual states in the US can choose to have minimum wages higher than the federal one, and almost 30 U.S. states do.

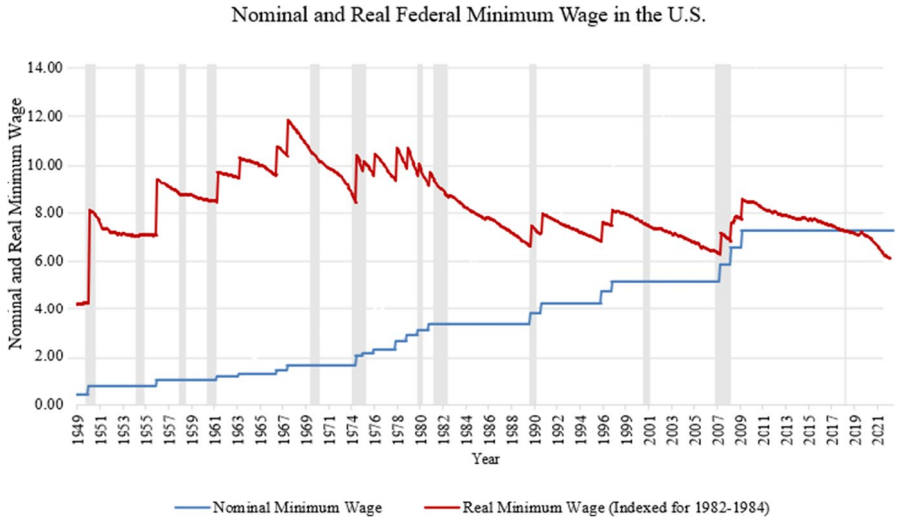

**Fig. 4** The federal minimum wage in the U.S. (nominal wage vs. real wage). Source of data: Bureau of Labor Statistics (BLS); U.S. Department of Labor–FRED Dataset. Note: Shaded areas indicate U.S. recessions

As shown in Table 1, the minimum wage in the US has been increased 20 times between 1949 and 2022, and each time, the newly set minimum wage has remained effective, on average, for a duration slightly exceeding three years and a half. Although the minimum wage has been increased so frequently, admittedly, there are no clear-cut criteria to use for making decisions about "when" the minimum wage should be increased, nor about the extent to which it should be increased. The focus of this paper is on the former (i.e., the timing of increase) and not the latter (i.e., the size of increase).[13] Unlike other economic variables such as Social Security (SS) payments and Cost of Living Adjustments (COLAs), which are indexed for inflation and thereby automatically increase as the Consumer Price Index (CPI) increases to compensate for inflation,[14] the minimum

---

[13] . In addition to the timing of increasing the minimum wage, another pertinent inquiry to consider is the extent to which the minimum wage should be raised. The latter matter lies outside the purview of this paper, and thus, it will not be addressed here. Nonetheless, in order to complete the discussion on the optimal timing of implementing this policy, a succinct response to the latter inquiry for the sake of this paper would be that the extent of the increase should align with the targeted real federal minimum wage established by the policy maker, whatever it is, which may be determined by considering factors such as the inflation rate, the poverty line, the free-market equilibrium wage for unskilled worker and youths, and other relevant considerations.

[14] A variable is called indexed for inflation when it is increased or decreased automatically according to the changes in the CPI to account for changes in prices due to inflation.

Table 1 The historical averages and characteristics of the minimum wage changes in the U.S. from 1949 until 2022

| Criterion | Value |
| --- | --- |
| Average Nominal Min Wage (based on counts) (Nominal $) | $3.03 |
| Average Nominal Min Wage (adjusted for time remaining effective) (Nominal $) | $3.65 |
| Lowest Nominal Min Wage in the Studied Period (which was in place in 1949) (Nominal $)[a] | $0.40 |
| Highest Nominal Min Wage in the Studied Period (which was set in 2009) (Nominal $) | $7.25 |
| Average Real Min Wage (based on average of purchasing power of $ between 1982–84) | $8.20 |
| Lowest Real Min Wage in the Studied Period (which was in effect in 1949/Jan) (Real $) | $4.21 |
| Highest Real Min Wage in the Studied Period (which was in effect in 1968/Feb) (Real $) | $11.83 |
| Number of Distinctive Nominal Wages | 21 |
| Number of Changes in Nominal Wage | 20 |
| Average Duration of Time for Each Minimum Wage to Remain Effective (years) | 3.62 |

[a]It is important to note that this is not the lowest nominal wage in U.S. history, but rather the lowest within the specific period under study in this paper. The lowest nominal wage in the U.S. has been the first minimum wage ever set in the U.S., which was $0.25 per hour established in 1938 (whose purchasing power was equivalent to almost $5.20 in terms of 2022 dollars)

Source of Data: Bureau of Labor Statistics (BLS); U.S. Department of Labor–FRED Dataset

wage requires legislative action from Congress to increase and is not tied to any price indices such as the CPI. This may be a deliberate approach taken to avoid the potential risk of wage-price spirals and continuous cycles of inflation as a result of that. However, if indexation is not deemed to be a suitable mechanism to decide about the minimum wage in a meaningful and systematical manner, there is a need for other clear, specified, and well-defined decision rules to replace the current ad-hoc practices and procedures when making decisions about the optimal timing of increasing the minimum wage. This section aims to provide such a decision rule that aligns with the dual policy objective of promoting efficiency and equality simultaneously.

While examining the historical trend of the real federal minimum wage in the US, it seems that the real federal minimum wage in the US has had an unwritten, unofficial lower bound of nearly $8 in the 1960s-1970s and a lower bound of almost $7 in the 1980s-2020s. However, despite these observations, there have still been no criteria officially defined to determine "when" the federal minimum wage needs to be increased. Instead, the path of the minimum wage has seemingly followed more of a random-walk process rather than being driven by clear policy considerations. As noted earlier, the minimum wage as an economic variable is not indexed for inflation and is increased at the discretion of Congress and signed into law by presidents. As such, the unique characteristics of the minimum wage as an economic variable provides an opportunity for it to serve as a policy instrument to not only stabilize the economy and promote efficiency thereby, but also to promote equality, in the way that is further elaborated in the following discussion.

A sensible temporal criterion for determining the optimal timing of increasing the minimum wage in such a way that it can promote both efficiency and equality simultaneously can be derived from the relationship between the natural rate of unemployment[15] and the actual rate of unemployment. In fact, an optimal timing for increasing the minimum wage can be when the actual rate of unemployment falls and remains for a while below the natural rate of unemployment. During such circumstances, the labor market experiences a shortage of available workers, and businesses are in need of additional workers. By increasing the minimum wage then, more individuals are incentivized to transition from the pool of those unwilling to work and join the pool of labor force participants. Moreover, it encourages existing workers to extend their working hours and engage in overtime working.

These developments collectively contribute to moving the labor market closer to a new equilibrium, above which the new minimum wage must be placed to be effective. Consequently, this timing of adjusting the minimum wage facilitates the convergence of the AD in the economy towards the LRAS. This alignment allows the economy to operate more efficiently by striking a balance where resources are neither underutilized nor overutilized, and thereby mitigates the possibility of recessionary or inflationary periods.[16] This way, both efficiency and equality are promoted, as the economy is stabilized and minimum-wage earners (who are often relatively disadvantaged) get to earn higher incomes.

Figure 5 depicts the actual rate of unemployment rate (which is the fluctuating line in blue) and the natural rate of unemployment (which is the smoother line in red).

In a sense, this paper proposes that there may be an alternative way to the current ad hoc decision-making regarding the timing of increasing the minimum wage in such a way that not only increases equality but also promotes efficiency. As introduced above, the proposal is that, if the objective of economic policy is to promote both efficiency and equality simultaneously, one avenue to achieve this dual objective is raising the minimum wage during periods when the actual rate

---

[15] The natural rate of unemployment is closely related to other macroeconomic concepts such as LRAS. The natural rate of unemployment is the level of unemployment that exists even when the economy is in normal situations and is operating at its full potential, and there is no cyclical (demand-related) unemployment. It represents the rate of unemployment that is consistent with stable inflation and reflects structural and frictional factors in the labor market. In short, the natural rate of unemployment is the rate at which the labor market reaches its equilibrium. The natural rate of unemployment is an important component of the Long-Run Aggregate Supply (LRAS) curve. The LRAS curve represents the level of output that can be sustained in the long run when all resources, including labor, are fully utilized. At the natural rate of unemployment, the economy is operating at its potential output level, which corresponds to the level of output determined by the LRAS curve.

[16] One may argue that increasing the minimum wage would increase the structural rate of unemployment in the economy. However, this argument holds less weight when the overall unemployment rate is low. This is because, in that situation, demand for labor becomes very inelastic, meaning that the size of "outsiders" will become negligible. Thus, in such a scenario, raising the min wage will cause only a very minimal increase in the structural unemployment. Additionally, when the actual rate is below the natural rate of employment, as described above, the cyclical rate of unemployment is negative. In such a situation, even a slight increase in the structural rate of unemployment will only fill the void left by the negative rate of cyclical unemployment. This is because the economy is already over-employed, and businesses need more workers, and it is illegal to sign a contract below the already existing minimum wage, which is of no interest to the potential workers, since if it was, they had already joined the labor market. In sum, while concerns about the potential impact of raising the minimum wage on structural unemployment exist, they are less significant in a low-unemployment-rate environment with relatively inelastic labor demand.

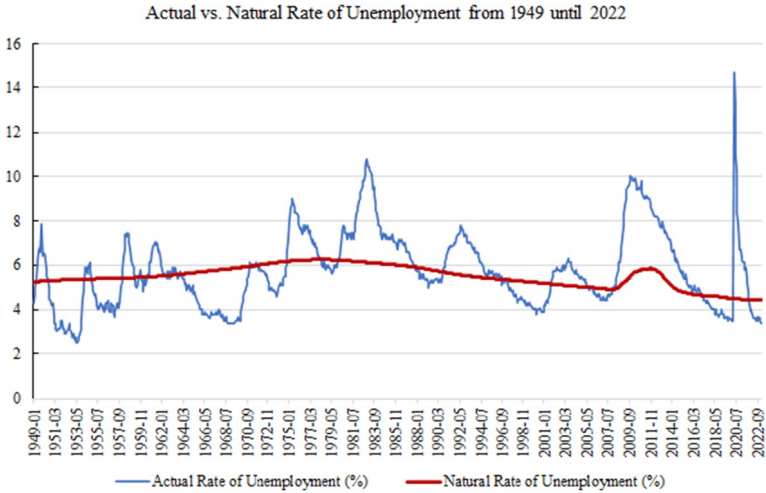

**Fig. 5** Actual unemployment rate and natural unemployment rate. Source of data: Bureau of Labor Statistics (BLS); U.S. Department of Labor–FRED Dataset. Note: Shaded areas indicate U.S. recessions

of unemployment falls and remains below the natural rate of unemployment for a while below the natural rate of unemployment. There are dozens of such instances in Fig. 5. For example, see four notable examples in periods between 1950–1953, 1964–1970, 1996–2001, and 2017–2019. In such a scenario, the negative effects of minimum wage laws on efficiency (e.g., an increased structural unemployment) can be mitigated and even offset by fulfilling negative cyclical unemployment rates in these periods. In this case, those negative effects are compensated by more efficiency gains from getting minimum-wage workers to businesses that need them, allowing for the possibility of achieving both efficiency and equality simultaneously.

The gap between the natural rate of unemployment and the actual rate of unemployment is equal to the cyclical unemployment rate. Cyclical unemployment refers to the portion of unemployment that is caused by fluctuations in business cycles or economic conditions. When the actual rate of unemployment is higher than the natural rate, it indicates an economic downturn or recession, and the difference between the two rates represents the cyclical unemployment rate. Conversely, when the actual rate is lower than the natural rate, it suggests an economic over-expansion and possibly inflation, and the gap between the said two rates represents a negative cyclical unemployment rate, indicating an overheating economy. Putting differently the goal of the present paper, this paper argues that, if the objective of economic policy is to promote efficiency and equality at the same time, the optimal timing of raising the minimum wage is at times when the cyclical rate of unemployment is negative for a period of time, and the labor market has not been able to clear and get back to its new equilibrium given the low minimum wage set in the economy. Figure 6 depicts the cyclical rate of unemployment rate (i.e., the gap) versus the time trend of minimum wage increases.

As shown in this figure, apparently, there are no clear temporal relationships between the cyclical rate of unemployment and the time trend of minimum wage

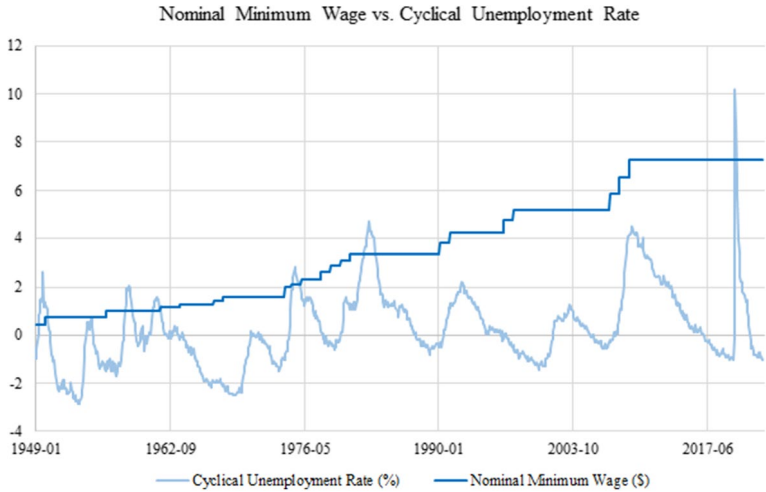

**Fig. 6** Nominal minimum wage vs. cyclical unemployment rate. Source of data: Bureau of Labor Statistics (BLS); U.S. Department of Labor–FRED Dataset

increases (as, in some cases, the minimum wage has been increased when the cyclical rate has been negative, and in some other cases, it has been increased when the cyclical rate has been positive). However, this paper argues that current ad-hoc practices and procedures for determining the optimal timing of minimum wage increases should be replaced with specified, well-defined decision rules. One such rule, as discussed in this paper, is to increase the minimum wage when the cyclical rate of unemployment is negative. It seems that the minimum wage has historically been increased with a primary focus on equality only. As proposed in this paper, however, it can also serve as a policy instrument for economic stabilization, helping to cool down an overheating economy, bringing the actual rate of unemployment rate closer to its natural rate, and bringing back the AD and LRAS closer to each other (all of which promote efficiency), while simultaneously promoting equality.

It is also important to note that this way of utilizing the minimum wage to implement stabilization policy is likely to encounter less resistance from political parties that routinely prioritize efficiency and those who advocate for businesses and believe that society as a whole will be better off if businesses are supported. This is because, under the circumstances described above, businesses are already facing labor shortages, and an increase in the minimum wage would incentivize more individuals to enter the labor market.[17]

---

[17] In a sense, it can be said that the political environment in many countries exhibit a dichotomy of efficiency-oriented parties and equality-oriented parties. The former group tends to advocate for labor rights and strives for social homogeneity as an ideal norm that should exist, while the latter group emphasizes business interests and accepts social hierarchy as an inevitable reality. Notable examples of the efficiency-oriented parties include the Republican Party in the US and the Conservative Party in the UK. Examples of the equality-oriented parties include the Democratic Party in the US and the Labour Party in the UK. These parties often represent contrasting ideologies and policy priorities, reflecting differing views on the balance between equality and efficiency in governance.

Hypothetically, if the proposal made in this paper had been followed in U.S. history as a decision rule to determine the optimal timing of increasing the minimum wage, it would have resulted in a different trajectory for the minimum wage (assuming the same starting point of $0.40 and the same ending point of $7.25 for the minimum wage during 1949–2022). Figure 7 illustrates two such possible paths, where, in one case, the middle of each period with negative cyclical unemployment rate is considered the optimal point in time (referred to as the Midpoint Method), and in the other case, the local minimum point in each period of negative rate of cyclical unemployment is considered the optimal timing point in time (referred to as the Local-Min Method). Either method has been implemented by two types of averages (arithmetic and geometric) to reach from the starting point ($0.40) to the ending point ($7.25). Additionally, the last two diagrams in the lower row demonstrate the proposed paths alongside the cyclical rate of unemployment, so the correspondence between the two can be viewed easily, where the minimum wage is raised whenever the cyclical rate of unemployment becomes negative and remains so for a while.

As shown in Fig. 7, both trajectory paths proposed for the minimum wage (those derived from the Midpoint method and the Local-Min method) share a common starting point of $0.40 and ending point of $7.25 for the minimum wage during 1949–2022. The reason for this common range is because the focus of this paper is not on the optimal "magnitude" of the minimum-wage increase, but instead on the optimal "timing" of the minimum-wage increase in such a way that it serves both efficiency and equality at the same time. Therefore, the overall size of growth from the beginning to the end point of the trajectories is kept constant, so the focus can be limited to the "timing" of the policy, which is the main focus of the proposal.[18]

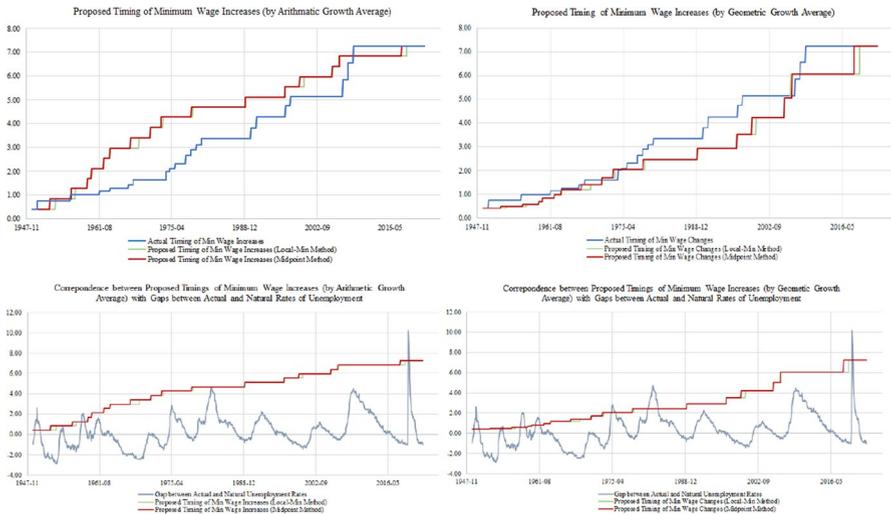

**Fig. 7** Actual trend of changing the nominal minimum wage with ad-hoc timing vs. proposed time trend for increasing the nominal minimum wage to achieve the dual objective of promoting efficiency and equality simultaneously. Source of data: Bureau of Labor Statistics (BLS); U.S. Department of Labor–FRED Dataset

As for determining the timing of raising, two methods are used called the Midpoint method and the Local-Min method.[19] The resulting trajectory paths from these two methods coincide and overlay in most cases, and the differences in the proposed paths are very minimal, as shown in the above figure. This high degree of coincidence and overlap occurs primarily due to the somewhat symmetrical nature of the time trend of actual unemployment rate when it falls below the actual rate of unemployment, resulting in a roughly V-shaped movement of a fall and rise, in which case both the local minimum points of the series overlap with their middle points or fall in close proximity of each other.

Additionally, to grow the minimum wage at the starting point to the minimum wage at the end point, two types of averages (i.e., arithmetic and geometric) have been utilized. Unsurprisingly, the arithmetic average results in faster growth in the minimum wage in the initial periods, while the geometric average does so in the final periods. As the variable of interest in plotting the trajectory paths is the minimum wage, which is a variable that should always be assessed in relation to purchasing power, the utilization of the geometric average is logically more suitable.

The two diagrams in the lower row provide a visual representation of the alignment and coincidence between the proposed optimal timing of raising the minimum wage with the periods of negative cyclical rate of unemployment under each regime of growth (i.e., arithmetic and geometric). As it can be seen easily in these diagrams, by following the optimal timing proposed in this paper, the minimum wage is increased when the cyclical rate of unemployment rate turns negative, and this specific timing of increase results in the promotion of both efficiency and equality concurrently. This stands in contrast to the ad-hoc and theoretically unfounded manner in which the minimum wage has traditionally been increased in U.S. history.

---

[18] As noted earlier, this paper does not attend to the question of "to what extent" the minimum wage should have been increased each time in the US. In all honesty, this question has often been attended to in a subjective, ad-hoc manner from an economic standpoint, and the minimum wage has never been treated as a policy variable, whose choice should follow certain well-defined decision rules. Typically, the process of adjusting the minimum wage entails deliberations and negotiations among federal lawmakers, and decisions about increasing it (both in terms of the size and timing of the increase) are handled mostly subjectively and by votes, rather than objectively and by rules. To avoid engaging in that subjective discussion, the paper maintains the same overall increase in the minimum wage during the study period, but it makes differing recommendations as to "when" the average increases should have been implemented to promote both efficiency and equality. The proposed timing of increases either precede or succeed the actual timing of increases enacted in the real world by the legislative actions, meaning that the increases should have been implemented either earlier or later in order to effectively achieve the dual objective of enhancing both efficiency and equality.

[19] These are only two illustrative examples showing how the timing of increases in the minimum wage can be adjusted to serve both equality and efficiency at the same time. Needless to say, these examples are not exhaustive, as there are other potential approaches to making this timing decision. A proper decision rule could be based on the typical anatomy and usual patterns of the statics and dynamics of business cycles (including four stages of expansions, peaks, recessions, and troughs) in the U.S. Alternatively, it could be based on setting a pre-determined downward deviation from the natural rate of unemployment. In this particular study, the criteria used as the decision rule are local minima and midpoints. As the analysis in this paper is being conducted *ex-post*, it allows for using such decision rules since the real-world phenomenon is being studied after the fact.

To be candid, this proposed way of timing minimum-wage increases serves economic efficiency (in terms of economic stabilization) more than income equality. This is because increasing the minimum wage at any time is most likely to promote equality, but the timing proposed here helps in mitigating its efficiency losses within the labor market and promote macroeconomic stability, thereby generating some efficiency gains in the macroeconomic environment. When moving from the same starting point to the same ending point earlier or later, almost the same amount of equality is achieved sooner or later. However, increased efficiency is the net gain of the proposed timing for the implementation of minimum wage increases, since utilizing the minimum wage as a policy instrument helps in stabilizing the economy with an eye to promoting equality. That being said, it is essential to consider a few additional points and caveats that follow.

One notable limitation of using the minimum wage as a policy instrument to achieve economic stability, efficiency, and equality simultaneously is that it is a policy instrument that can be used only in one direction (i.e., it can only be increased, and not decreased). From a political-economy perspective, reducing the minimum wage would be politically unattractive and unfeasible, since it poses significant challenges and risks for politicians, given the social tensions and political costs associated with such a decision for lawmakers and policymakers. In that sense, the minimum wage can be adjusted only in one direction (i.e., upward) when being used to stabilize the economy. This stands in contrast with the use of other policy instruments that are commonly used for stabilization, such as interest rates, which can be increased or decreased (either through Open Market Operations (OMOs), the discount rate, reserve requirements, or Interest on Excess Reserves (IOER) rates), without receiving significant opposition or controversy. Therefore, realistically, governments are unlikely to lower the nominal minimum wage, but they may choose to maintain it at its then-current level for an extended period of time, allowing inflation to erode its purchasing power and effectually reduce the real minimum wage over time.

It is also important to discuss further the reach and scope of the proposed policy in this paper. As mentioned earlier, the minimum wage as an economic variable is not indexed for inflation apparently for one primary reason, which is to prevent a wage-price spiral from occurring in the economy. However, formulating a well-defined decision rule for the optimal timing of its increases will not pose much threat due to the potential emergence of wage-price spiral. This is because minimum wage earners constitute a very small portion of the workforce in the US. According to BLS (2021), less than 2% of all hourly paid workers are federal minimum-wage earners. In fact, most workers are non-minimum-wage earners. Although increasing minimum wage can have a propagative effect on other market wage rates existing in its proximity (as it may be used as a benchmark to set other hourly wages in its upward neighborhood), all these groups that are paid wages close to the minimum wage are essentially disadvantaged groups, and raising the minimum wage will eventually increase equality in society. Thereby, when the actual rate of unemployment is lower than the natural rate of unemployment,[20] min wage can be used as a policy variable to promote both equality and efficiency in the economy at the same time. Thus, all these keep the proposal in line with one aspect of the dual objective which is promoting equality.

Additionally, one may argue that increasing the minimum wage can increase AD in an already booming economy, exacerbating the inflationary gap between AD and AS in the economy. However, this argument overlooks two important factors. Firstly, the minimum wage earners constitute less than 2% of the labor force, and not all the labor. Therefore, the effect of such an increase in AD is minimal. Secondly, the increase in the minimum wage has also a positive effect on AS in the economy by addressing labor shortages during periods of negative cyclical unemployment rates. These shortages typically turn to bottlenecks in the supply chain during such periods, which usually necessitate immediate actions and response rather than relying solely on market forces taking their course to naturally resolve the issue in the longer run. Given the fact that the economy is experiencing a labor shortage during these periods, an increase in the minimum wage would considerably help in increasing labor supply in the labor market, contributing significantly to a larger AS in the macroeconomy in the way outlined earlier. Given these two facts, the effect of an increase in the minimum wage on AS is very likely to outweigh that on AD. Therefore, the overall effect of reasonable minimum-wage increases during times of negative cyclical unemployment is expected to increase AS most likely more than increase AD, closing the gap between the two, which ultimately contributes to economic stabilization, which is a facet of economic efficiency at the macroeconomic level.

From an economic perspective, the absence of minimum wage laws would be preferable in general due to the potential market inefficiencies they can introduce in the labor market. However, from a political-economy point of view and in light of the prevalence of minimum wage laws in many countries nowadays, minimum wage laws have become a response to social pressures asking for them and a consideration of the broader political economy. This paper argues that if such laws are to exist, they should follow the dynamics and the timing proposed herein in order to contribute positively to economic stabilization by expediting the process of labor market clearing, and increasing the minimum wage when the old, in-place minimum wage becomes slack and unbinding (resulting from a rightward shift in the labor demand curve), and thereby increase efficiency by a dynamic move closer to a new equilibrium in the market while promoting equality.

A final caveat to consider is that the usage of the minimum wage as a policy instrument must be decided about with an eye to fiscal policy instruments (taxes and spending), and monetary policy instruments (interest rates, OMOs, etcetera). The interactions between these three sets of instruments must be given careful consideration. In a sense, it can be said that the Federal Reserve, in its pursuit of handling the short-run tradeoff between inflation and unemployment optimally, employs monetary policy by using interest rates as an instrument (along with its underlying tools of conducting monetary policy such as OMOs). On the other hand, the Congress and government are the primary institutions in charge of the optimal handling of efficiency-equality tradeoff, for which they can use the minimum wage as a policy instrument in addition to their traditional fiscal-policy instruments such as taxes and government spending. Therefore, it is essential to harmonize all these policy goals and instruments in order to ensure a high degree of synergy among various economic policy designs and efforts in the broader landscape of the economy. Coordinated and aligned policies can enhance the overall effectiveness of these policies and improve their collective impact on economic stability, growth, and equality, which ultimately can lead to a more cohesive, balanced, and stable economic system.

---

[20] More accurately, the implementation of this policy should be appropriate when the actual rate of unemployment among "unskilled" workers is lower than the natural rate for this particular group. The rationale behind this lies in the fact that unskilled workers constitute the primary target group affected by the minimum wage, which makes them the most relevant group to consider in the design and implementation of this policy. Therefore, an implicit assumption here is that the actual and natural rate of unemployment among unskilled workers follows the same trends as those of the labor force as a whole.

### 2.4 Motivating social cooperation in the case of externalities: equality and efficiency in the usage of natural resources

Externalities—also known as spillovers—refer to the unintended consequences of economic activities that affect third parties who are not directly involved in the transactions of an economic activity. Externalities can be positive or negative, depending on whether they generate a beneficial impact or an adverse impact on bystanders. This section deals primarily with negative externalities and particularly with the parable of the Tragedy of the Commons. The Tragedy of the Commons explains the concept of overexploitation of shared resources and the potential consequences of unregulated self-interest where short-run individual incentives do not align with long-run social incentives. By exploring the dynamics of strategic interactions between self-interested individuals and potential roles for the government to intervene as a regulator, policy-maker, or owner, it can be better understood how society can address the challenges that are typically posed by the Tragedy of the Commons, and in particular, how social institutions can aim for promoting efficiency and equality simultaneously under such circumstances.

The Tragedy of the Commons is a well-known concept in game theory that describes a situation where multiple individuals, acting independently and rationally, deplete a shared resource to their own detriment. Negative externalities occur when the actions of one individual or group impose costs on others without any compensation in return, and others may reciprocate by imposing similar costs. In that sense, the Tragedy of the Commons is a prime illustration of negative externalities in an interactive setting. In fact, it arises when multiple self-interested individuals exploit a common resource beyond its sustainable, carrying capacity, which ultimately results in its depletion or degradation that affects all members of society. In addressing the Tragedy of the Commons, game theory provides insights into the strategic interactions between self-interested individuals by analyzing the players' choices, payoffs, and incentives to identify natural equilibria that may arise, such as the Nash equilibrium, which represents a stable state where no player can improve their outcome by unilaterally deviating from that state.

In the context of a common pasture, for example, each herder has an incentive to graze more livestock for personal gain. However, when every herder pursues this strategy under the erroneous assumption that other herders would not think similarly (concluding that the common resource will remain in existence), the pasture becomes overgrazed. This results in reduced productivity and long-term harms for all herders, leading to lower efficiency compared to when all herders use the resource in a sustainable manner. This situation highlights how rational decision-making based on self-interest can lead to suboptimal outcomes for society as a whole in certain circumstances.

As an illustrative example, consider a scenario where two herders graze their livestock on a public pasture. They independently choose between using the common resource "lightly" (at a sustainable rate) or "heavily" (at a destructive rate). When they graze a small number of livestock (i.e., Light), the resource can be sustained indefinitely and can be used forever. However, when they exceed the tipping point of the pasture and graze a large number of livestock (i.e., Heavy), the resource will soon be depleted and cannot be sustained. The resulting strategic interactions and payoffs can be modeled as a non-cooperative, symmetric, simultaneous-move, single-shot game. The following payoff matrix (Fig. 8) provides an example representation of their outcomes with the payoffs being the benefits from sustainable cooperation in the long run.

In this interactive context and strategic framework, each of the herders has a dominant strategy of choosing "Heavy" and graze a large number of their livestock. This results in a Nash Equilibrium of NE = {Heavy, Heavy}, which delivers a modest payoff of 30 units to each herder and a collective outcome of 60 units for this two-member society. Undoubtedly, it stands as the worst collective outcome among the four possible outcomes, as it delivers the lowest collective payoff to the two herders among all possible outcomes in this strategic game, due to failing to align with the long-term interests of both herders. In fact, although the Best Social Outcome (BSO) is delivered when the two herders both choose to graze lightly to end up with

**Fig. 8** Game-theoretic representation (payoff matrix) of the tragedy of the commons

|  | | Herder 2 | |
|---|---|---|---|
|  | | Light | Heavy |
| Herder 1 | Light | 50, 50 | 60, 20 |
| | Heavy | 20, 60 | 30, 30 |

50 units of benefits for each and a remarkable social benefit of 100 units. The reason why the parties still end up with the Nash equilibrium naturally is because they think strategically, pursue self-interest, make privately rational decisions independently of each other and under uncertainty about the other player's chosen action, and thereby, choose to graze heavily knowing that there is no guarantee that the other party does not do so. As a result, they both end up with the worst possible collective outcome. However, the best social outcome among the four possible outcomes of this strategic situation is BSO = {Light, Light}, as it delivers the highest amount of output from the scarce resource, indicating the highest level of efficiency. Additionally, in terms of equality, the difference between the two outcomes in each cell under each possible strategy set can be regarded as a measure of inequality under each possible outcome, as it represents the gap between the two payouts to the two players. Accordingly, the two outcomes of {Heavy, Heavy} (i.e., the Nash Equilibrium) and {Light, Light} (the best collective outcome) yield the same level of equality, while {Light, Light} embodies the highest level of efficiency, too.

Under the Nash equilibrium, each player, driven by self-interest, acts in a way that maximizes their individual benefit. However, as shown above, the Nash equilibrium is far from social optimum within the context of the Tragedy of the Commons. It leads to overexploitation of the resource, results in negative externalities, and diminishes output and welfare for all participants, which reduces overall efficiency. This arises due to the fact that each individual herder is driven by an erroneous belief in an equal opportunity in overgrazing, which leaves society with less efficiency when the whole community thinks mistakenly that their over-usage of the pasture will not result in the depletion of the resource. This situation may also arise when participants believe that refraining from overgrazing would prompt other herders to engage in overgrazing first, which leads them to conclude that it is more advantageous to overgraze preemptively. In a sense, the direction of these individual incentives is unintentionally towards a race to the bottom for the society as a whole.

Although the best social outcome is achieved under the cooperative strategy set of {Light, Light}, where both herders choose "Light," the idea of self-interest prevails eventually, as they are most likely to put their own short-run benefits first when they cannot be certain about the long-run cooperation of the other player. In this case of negative externalities, which constitutes a systemic market failure, the parties do not naturally end up with the best social outcome when they pursue self-interest. Hence, it becomes imperative to explore effective solutions that can steer the two parties away from the Nash equilibrium outcome, which emerges as a result of adhering to their natural incentives, towards the best social outcome. Any such solution will promote efficiency and uphold equality simultaneously, as explained in what follows.

There are various courses of action and policies that can make this two-member society play socially optimal strategy plans and guide it towards the socially best outcome. For example, the government can use public solutions such as command-and-control policies (e.g., setting limits and using technology to monitor compliance), market-based policies such as corrective taxes (also known as Pigouvian taxes), subsidies, and tradable permits. These solutions are presented here in the order of their related monitoring costs, from most to least expensive. Alternatively, this problem can be addressed through non-governmental, private entities, so that the private parties can solve the problem on their own without government intervening as an intermediary. Examples of private solutions to this problem include relying on moral codes and social sanctions, assigning property rights, and utilizing contracts between market participants and bystanders (as advocated by Ronald Coase (1960) through the Coase Theorem, which can be applied as long as the required conditions for the Coase theorem holds). In these cases, there will be no monitoring or intermediary costs imposed by the government, so the problem is resolved in a more efficient way.

To address the Tragedy of the Commons, which is fundamentally a systemic error, the involvement of either government intervention or private mechanisms becomes crucial. The aforementioned policies, regulations, and mechanisms can be implemented and utilized to internalize the negative externalities and guide the behavior of self-interested individuals towards more socially desirable outcomes. By using these approaches and instruments to make the players shift from NE = {Heavy, Heavy} to BSO = {Light, Light}, participants are incentivized to consider the long-term consequences of their actions, make sustainable choices, and limit the extraction or utilization of the resource to ensure its preservation for themselves as well as their future generations. These policies aim to promote efficiency and equality by considering individual incentives and collective interests.

When the two players are guided to play the best social outcome as demonstrated in the example above, they not only achieve the same level of equality in the consumption of the resource (as opposed to an equality in the destruction of the resource, which was the case when in the Nash equilibrium), but they also attain a higher (sustainable) level of efficiency in the usage of the resource. This sustainability has two important aspects. First, it allows for the maximal usage of the scarce resource in a sustainable manner. Secondly, it guarantees the preservation of the natural resource for future generations, which can be interpreted as a promotion in intra-generational equality. Therefore, not only do these methods of regulating the resource preserve the equality level at a sustainable level of the usage of the resource within the same generation (as it was the case with absolute equality in the illustrative example provided above), but it also promotes intra-generational equality in having access to natural resources in the future.

In sum, employing the solutions discussed in this section enables societies to encourage and incentivize a type of social cooperation that enhances efficiency (by mitigating negative externalities and ensuring the sustainable use of shared resources), promotes equal opportunity, preserves full equality in payoffs earned, and also fosters intra-generational equality at the same time. Therefore, the tradeoff between efficiency and equality can be avoided in such settings through social cooperation.

### 2.5 Encouraging charitable engagement for social impact in the economy: maximizing the economic pie for the poor and maximizing the utility pie for the rich

Most government policies that are aimed at promoting income equality involve redistributive measures, such as levying higher taxes on affluent individuals and redistributing the collected proceeds to support social welfare programs targeted at serving the less fortunate. While the underlying objective of these governmental policies is to address income inequality and provide assistance to vulnerable groups, it is crucial to conduct a comprehensive analysis of the potential consequences they may entail for both the affluent and the less fortunate. This section seeks to present a rational argument to highlight the potential detrimental effects of imposing heavy taxation on the wealthy's incentives for work and investment and consequently on the poor's productivity and well-being.

In order to comprehensively assess the effects of redistributive measures, it is important to acknowledge that the wealthy constitute a sizable proportion of entrepreneurs and business owners in most societies. Therefore, they are the ones who predominantly assume the role of decision-makers about the extent of engagement in economic activity and the scale of production while they are pursuing self-interest and are driven by the profit motive. The imposition of high tax rates on the wealthy can undermine their motivation to engage diligently in economic activity and to participate in entrepreneurial efforts. When the wealthy witness that a substantial portion of their income from such efforts after the fact will be appropriated through taxation, they experience a sense of discouragement when considering their work-leisure choice problem and lose the motivation to take risks and invest in new ventures.[21]

Since high tax rates significantly reduce the disposable proceeds of the wealthy's entrepreneurial endeavors, such taxes reduce their incentives as well as abilities to expand and grow their already existing businesses. Moreover, high tax rates reduce their incentives and capabilities to invest in new enterprises. Due to these weakened incentives and limited financial resources, the wealthy do not invest in business growth and new entrepreneurial ventures, which negatively influence the formation of capital (i.e., factories, machinery, and equipment) in the economy eventually. As a consequence, this leads to a reduction in capital intensity due to capital shallowing taking place in the economy, which in turn, results in a declined level of labor productivity and a decreased scale of economic output, which altogether ultimately reduce the size of overall economic efficiency.

It is crucial to consider the broader perspective that the relationship between labor and capital, as two factors of production (i.e., inputs), is intrinsically more of a complementary relationship than a substitutable one. This fact is contrary to the customary belief that arises from focusing solely on isolated instances of labor jobs being replaced by capital and technology in the short run due to innovation.

---

[21] Research has shown that excessive taxation can have negative effects on entrepreneurial endeavors and job creation. High tax rates can result in reducing incentives for individuals to engage in entrepreneurial activities (as they may perceive the after-tax rewards to be insufficient), decreasing business start-ups and innovation, decreasing self-employed rates, increasing the administrative burden and compliance costs associated with complex tax systems (diverting resources away from productive activities), and influencing the location decisions of entrepreneurs (causing outflows of entrepreneurial talent), all of which can lead to lowering employment rates (Feld & Kirchgässner, 2003; Rathelot & Sillard 2008; Djankov et al. 2010; Rohlin et al. 2014; and Darnihamedani et al. 2018). These indicate that policies targeting the wealthy can harm the job prospects of the less privileged. When the rich are discouraged from expanding their businesses and investing in new ventures, the pool of available jobs shrinks, which leaves the poor with fewer job opportunities to escape poverty and improve their living standards.

These instances are eventually beneficial improvements in an economy overall and are caused by the process of creative destruction within a dynamic economy. However, they do not convey the reality of the relationship between labor and capital as two factors of production in the long run. Economic history has proven the complimentary relationship between these two factors of production over centuries; as more capital (i.e., factories, machinery, and equipment) has been accumulated in the world, more jobs for labor have been created as well.

Understanding the long-run interdependence of labor and capital reveals that, if more jobs are to be created, more investment needs to be made, so that more capital goods can be constructed for labor to be able to work with them productively. In that sense, investments play a pivotal role in the creation of new employment opportunities. When the rich have weaker incentives to invest, there will be a ripple effect throughout the economy, which affects the availability of capital for businesses to grow and thrive. Thus, this shortage of investment impedes the ability of businesses to provide job opportunities for the poor, who often rely on job opportunities provided by businesses, and these taken together limit employment prospects and diminishes the chance of upward mobility for the poor. In sum, by taxing the wealthy heavily, the government risks dampening down these investments and limiting job prospects for the poor.

This is in fact the way that the challenges posed by imposing high tax rates on the rich eventually extend to the poor themselves, the very group that the policymaker aimed to help in the first place. In other words, while the goal of income redistribution is to provide assistance to the less fortunate, excessive reliance on taxing the wealthy can have unintended consequences for the intended beneficiaries (i.e., the poor). As an example channel which was introduced above, when the rich face weakened incentives to invest in enterprises and create jobs, the poor may experience limited employment opportunities and decreased upward mobility. Additionally, the shrinkage of the economic pie resulting from decreased productivity can lead to a decrease in overall wealth and resources available for redistribution, potentially undermining the effectiveness of and reducing the scope of social welfare programs. Furthermore, when the government mandates the participation of individuals in compassionate income transfers through high tax rates, it may foster a counterproductive sense of entitlement among the poor. Rather than cultivating self-reliance and empowerment, an excessive reliance on government assistance may create a dependency mindset that perpetuates vicious cycles of poverty. It is imperative to prioritize policies that empower individuals, foster self-reliance, and maximize economic efficiency, as these ultimately create a more prosperous society for all and more equitable society for those who choose to contribute actively to economic activity.

Additionally, taxing the rich heavily also reduces the surplus that can potentially be used towards charitable activities. It is important to notice that the necessary condition for the feasibility of charitable activities is the very existence of a surplus in the hands of the people in the economy, which usually is eliminated by the deadweight loss and inefficiencies of government intervention as an intermediary in the economy. The lack of economic incentives for the rich can reduce their philanthropic activities, thereby negatively impacting the availability of private charitable initiatives that may provide more targeted and effective support to those in need without government intervening as an inefficient intermediary (also known as a "leaky bucket" as referred to by Arthur Okun).

Although various forms of redistributive measures can promote equality to some extent, almost all such forms come at the expense of efficiency when compared to a counterfactual state of the world where these mandatory redistributive measures are not implemented (assuming all other factors remaining the same). However, an alternative approach to promoting income equality in such a way that it promotes equality without harming efficiency involves encouraging charitable activities through policies such as tax breaks, tax deductions, tax credits, and tax exemptions.[22] This approach enables the rich to contribute voluntarily to causes they consider important, without compromising their incentives to work, produce, and invest in the economy. Furthermore, when individuals have the freedom to direct their charitable contributions, it enables them to support causes and initiatives that align more with their specific interests, expertise, and moral preferences. By supporting the voluntary forms of philanthropy, the government can harness the compassion of individuals to address social issues while preserving the incentives necessary for fostering economic growth and preserving economic efficiency.

Furthermore, when the government mandates by force the participation in the compassionate income transfer of their choosing, the size of happiness and utility pie will inadvertently shrink in the economy. In contrast, when the government aims to promote equality by encouraging charity activities through policies based on tax incentives, which are essentially done voluntarily, then this alternative way of financing the need for helping for the poor will not undermine the incentives of the rich. Hence, it will not have the aforementioned negative effects on the rich and the poor. Accordingly, both economic pie and utility pie are maximized, both of which are important facets of efficiency in an economy.

In sum, while the objective of promoting equality for individuals who have experienced systemic disadvantages is noble and commendable, it is essential to consider the potential unintended consequences of policies that heavily tax the rich, especially those consequences that negatively influence the very same group of the disadvantaged. It is also imperative to recognize the synergistic relationship between labor and capital as two complementary inputs in the economy. Recognizing the complementary nature of labor and capital underscores the importance of maintaining a conducive environment that encourages investment. Overloading the wealthy with high taxes diminishes their incentives to engage and invest in economic activity, which leads to adverse effects on the well-being of the poor, too, eventually. Through alternative approaches such as encouraging charitable activities through tax incentives, the government can strike a balance between promoting income equality and preserving the incentives necessary for economic efficiency.

---

[22] Studies have shown that policies encouraging the engagement in charitable activities can have positive effects on the individuals' willingness to participate in such activities. For example, a study by Clotfelter (1985) found that tax incentives for charitable donations increased giving levels, leading to greater support for charitable organizations and social programs. Lin and Lo (2012) studied the effect of tax incentives on charitable contributions using censored quantile regression model, concluding that tax incentives promote charitable activities. Duquette (2016) has investigated the effect of tax incentives on charitable contributions, finding that a one percent increase in the tax cost of giving causes charitable receipts to fall by about four percent. The proposed approach to achieve the dual objective of promoting equality and efficiency that was introduced in this section not only allows individuals to exercise their personal values and moral priorities but also fosters a sense of community and social responsibility.

## 3 Conclusion and summary

Economic efficiency and income equality are often thought of as two competing goals in the realm of economic policy. The idea is that policies that promote economic efficiency, such as low taxes and free markets, tend to result in income inequality, while policies that promote income equality, such as progressive taxation and redistribution of wealth, tend to come at the expense of economic efficiency. However, this tradeoff relationship does not have to always exist, as there are avenues through which economic efficiency and income equality can be promoted simultaneously. This paper identifies five specific circumstances where this tradeoff can be effectively avoided and illustrates why and how the tradeoff between these two contradictory societal goals can be cleverly averted under the mentioned circumstances. Additionally, it discusses the policy implications arising from these findings, emphasizing the importance of informed decision-making in balancing these societal goals.

The first circumstance in which the efficiency-equality tradeoff can be avoided is when aggregate demand (AD) falls short of aggregate supply (AS) in the economy. In that situation, the government has the option to stimulate AD by targeting individuals with the highest Marginal Propensity to Consume (MPC) in their expansionary fiscal policy. This approach will boost economic activity, stabilize the macroeconomic environment, promote efficiency (by addressing inefficiencies caused by underutilized resources during recessions), and promote equality (by designing tax cuts that benefit the poor relatively more) as well. In sum, when in a recessionary gap, targeting the poor in the design of an expansionary fiscal policy would help more effectually the economy in returning it back to the potential output trend, which results in enhancing both efficiency and equality at the same time.

The second avenue through with the government can promote efficiency and equality simultaneously is by providing equal opportunities and implementing empowerment initiatives (such as student loan programs) for the needy in order to enable them to become economically independent members of society instead of them imposing a significant financial burden on the public sector's budget in the future if they are left unsupported today. Such initiatives may seem to be a pressure on the public sector's budget in the short run, but they definitely come with long-run gains for society as a whole and generate considerable cost-savings for the public sector's budget in the long run, when compared to the counterfactual. A rational, marginal cost–benefit analysis is conducted in this paper for such a program from the viewpoint of the government's public budget, showing that the NPV of the benefits generated by providing such programs is very likely greater than the NPV of the public resources allocated towards such an empowerment program. This type of empowerment program serves as an exemplary effort that archives not only the dual objectives of enhancing income equality and economic efficiency, but also provide equal opportunities and generates positive externalities. As such, the implementation of such empowerment initiatives is not only of instrumental value but also of intrinsic value.

The third channel through which the big tradeoff can be averted is by optimizing the timing of raising the minimum wage in the economy. The paper proposes that there could be an alternative way to the current ad-hoc decision-making procedure regarding the timing of increasing the minimum wage in such a way that it not only increases equality but also promotes efficiency. This paper argues that this dual objective can be achieved if the minimum wage is raised when the actual rate of unemployment is below the natural rate of unemployment. This timing of raising the minimum wage contributes positively to economic stabilization, and thereby increases efficiency by a dynamic move closer to a new equilibrium in the market while promoting equality. In such a scenario, the negative effects of minimum wage laws on efficiency (such as an increased structural unemployment) can be mitigated and even offset by fulfilling negative cyclical unemployment rates in these periods. In this case, those negative effects are compensated for by more efficiency gains from getting minimum-wage workers to businesses that need them, allowing for the achievement of the dual objective.

The fourth avenue through which the big tradeoff can be avoided is by motivating social cooperation in the case of negative externalities arisen in the usage of natural resources. The paper uses game-theoretic models and attends to the challenges that are typically posed by the Tragedy of the Commons, and in particular, how social institutions can aim for promoting efficiency and equality simultaneously under such circumstances. It is shown that, under the Nash equilibrium, each player, driven by self-interest, acts in a way that maximizes their own individual benefit, while the Nash equilibrium is far from socially optimal outcome within the context of the Tragedy of the Commons. In fact, the direction of individual incentives is unintentionally towards a race to the bottom for society as a whole under these circumstances. The paper introduces various courses of action and policies that can be used to encourage and incentivize a type of social cooperation that enhances efficiency (by mitigating negative externalities and ensuring the sustainable use of shared resources), promotes equal opportunity, preserves full equality in payoffs earned, and also fosters intra-generational equality at the same time.

The fifth channel proposed in this paper through which efficiency and equality can be promoted simultaneously is by encouraging charitable engagement for social impact in the economy, which can maximize the economic pie for the poor and maximize the utility pie for the rich. It is argued that most government policies that are aimed at promoting income equality involve redistributive measures, which typically generate a loss of efficiency primarily due to a shrinkage in the formation of capital. It is argued that the relationship between labor and capital, as two factors of production, is more of a complementary relationship. Therefore, if more jobs are to be created, more capital investment needs to be made, so that more capital goods can be constructed for labor to be able to work with them productively. The shortage of investment impedes the ability of businesses to provide job opportunities for the poor, who often rely on job opportunities provided by businesses, and these taken together limit employment prospects and diminishes the chance of upward mobility for the poor. In sum, by taxing the wealthy heavily, the government risks dampening down these investments and limiting job prospects for the poor, the very group that the policymaker aimed to help in the first place. However, an alternative approach to promoting income equality in such a way that it promotes equality without harming efficiency involves encouraging charitable activities through policies such as tax breaks, tax deductions, tax credits, and tax exemptions. This approach enables the rich to contribute voluntarily to causes they consider important, without compromising their incentives to work, produce, and invest in the economy. This way, the government can achieve the two goals of promoting income equality and preserving the incentives necessary for economic efficiency at the same time.[23]

Ultimately, although the efficiency-equality tradeoff is more or less an unavoidable tradeoff in many socio-economic settings, there are still certain circumstances, such as those outlined above, in which this tradeoff can be averted. A benevolent social planner should make every effort in order to deftly avert this tradeoff. This paper contributes to the literature on economic policy particularly in terms of exploring the pursuit of multi-purpose socio-economic policies. The identified avenues presented in this paper hold considerable potential to enable market-based economies to concurrently promote both efficiency and equality under the described situations and evade the significant social challenges posed by the so-called "Big Tradeoff."

---

[23] It is important to note that it is crucial to also consider the varying institutional frameworks and policy environments that may exist among advanced, emerging, and developing economies. While stabilization policies and laws such as the minimum wage law are nowadays implemented in many emerging and developing economies similarly to those in advanced economies, and given the fact that many countries may share similar institutional structures, it is still essential to recognize that each economy may face unique challenges and constraints. Such a heterogeneity in institutional foundations suggests that, while certain policy interventions may be effective in promoting both equity and efficiency in advanced economies, the same strategies may yield different outcomes in emerging and developing economies due to differences in institutional capacities, market structures, and levels of economic development. In light of these considerations, it is plausible to argue that while the proposed avenues may offer valuable insights and potential solutions, their implementation must be tailored to the specific context of each economy. Policymakers should take into account the unique socio-economic conditions and institutional capacities of their respective countries when considering the adoption of these strategies, thereby ensuring their effectiveness and relevance across diverse economic environments.

**Acknowledgements** I would like to extend my sincere gratitude to the participants and discussants at my presentations of this paper for their invaluable comments and suggestions on earlier drafts. I am also deeply grateful to the Wake Forest University Center for the Study of Capitalism for their support. The views expressed in this paper are my own and do not necessarily reflect those of other institutions. Additionally, I wish to thank the two anonymous reviewers and the editors for their insightful and constructive feedback, which greatly improved the quality of this work. Any remaining errors are entirely my responsibility.

**Funding** Open access funding provided by the Carolinas Consortium.

**Data Availability** The data supporting the findings of this study are available from the sources listed under the figures and tables within the paper, and also can be obtained from the author upon reasonable request.